\documentstyle[multicol,pra,aps,epsf,epsfig]{revtex}
\def \l{\lambda}
\def \dg{\dagger}
\def \rof{\rho_f^{(ss)}}
\begin{document}

\title{Aspects of nonlocality in atom-photon interactions in a cavity}

\author{A. S. Majumdar\footnote{archan@boson.bose.res.in}$^1$ and N. Nayak\footnote{On leave from S. N. Bose National Centre for Basic Sciences, Calcutta, India.\\~~~~~~~~ nayak@boson.bose.res.in,~nayak@atlantic.tamu.edu}$^2$}

\address{$^1$S. N. Bose National Centre for Basic Sciences, 
Block JD, 
Sector III, Salt Lake, Calcutta 700098, India}

\address{$^2$Department of Physics, Texas A{\&}M University, College Station, 
Texas 77843-4242, USA}

\maketitle


\begin{abstract}

We investigate a Bell-type inequality for probabilities of
detected atoms formulated using atom-photon interactions in a cavity. We
consider decoherence brought about by both atomic decay, as well as cavity
photon loss, and study its  quantitative action in diminishing the atom-field
and the resultant atom-atom secondary correlations. 
We show that the effects of decoherence
on nonlocality can be observed in a controlled manner in actual experiments
involving the micromaser and also the microlaser.

PACS number(s): 03.65.Bz, 42.50.Dv

\end{abstract}


\section{Introduction}

Nonlocality at the quantum level manifests itself in various kinds of phenomena.
The study of this so far, has been predominantly confined to the study of
interactions amongst similar kinds of particles, for example, photon-photon
interactions or the interaction of subatomic particles among themselves.
With the advance of technology over the last several years, it has now become
conceivable to investigate a new kind of nonlocality in a controllable fashion, 
viz., the nonlocality generated through the interaction of distinct entities, 
like atoms and photons inside a high quality cavity.

The mathematical framework for demonstrating the violation of local realism in
quantum mechanics was first provided by Bell\cite{bell} through his famous
inequalities. This work was subsequently  generalized and also 
extended to consider the
interaction of more than two particles\cite{review,chsh}. A different
kind of proof of nonlocality without the use of inequalities, also exists
\cite{hardy}.
Furthermore, it has
been shown that quantum nonlocality continues to persist even for the case
of a large number of particles, or large quantum numbers\cite{macro}. This
has raised certain questions regarding the issue of the macroscopic
or classical limit of quantum mechanics in examples where both the number of
particles, and their quantum number is made arbitrarily large\cite{larjn}.
The phenomenon of environment induced decoherence is of direct relevance
here. It would be interesting if decoherence could be experimentally
controlled and its effect on nonlocality be quantitatively monitored in
particular examples of study.

Experimental proposals of demonstrating nonlocality have mostly been concerned with
spin-$1/2$ particles\cite{proposals}, photons\cite{photon1,photon2}, or mesons\cite{mesons}.
In recent times, several schemes involving two-level Rydberg atoms have been 
proposed\cite{rydberg1,rydberg2}. In such schemes two-level rydberg atoms are considered in analogy
 to the spin system in Bell's original reasoning. The role of the polarization
axis of the Stern-Gerlach apparatus used for spin-$1/2$ systems is played
here by the phase of an auxiliary electromagnetic field.
The primary advantages of the experiments
using atoms, compared to those with photons or spin half particles are two-fold. First, the
realization of spacelike separation for Rydberg atoms is easier because of their smaller
velocities than photons or electrons. Secondly, the efficiencies of detectors used for
the former is much larger in general in comparison with the detectors used for elementary
particles. In addition, the interaction of large-sized atoms with the environment can
be significant enough to be monitored in certain cases. In fact, in the experimental
schemes which involve the interaction of Rydberg atoms with photons in a microwave
cavity, dissipation through the loss of cavity photons  always occurs. The effect of this
is manifested in the form of loss of coherence in the atom-photon interactions. Thus, this
is a natural arena to study the effects of decoherence on quantum nonlocality in a 
quantitative manner.

In this paper we  propose a realistic experiment to
test Bell's inequality for real micromasers/microlasers by taking
experimentally attainable data in the presence of both atomic decay
and cavity dissipation under the influence of their respective
reservoirs. The approach followed by us enables us to 
analyze both  micromaser as well
as  microlaser dynamics within the framework of a single formalism 
in presence of decoherence.
Our aim is to study the effects of decoherence on the
magnitude of violation of Bell's inequality in an experimentally
controllable fashion. As an interesting sidelight, we are also able
to demonstrate the effect of decoherence on multiparticle
correlations that are a natural outcrop of experimental schemes
using microwave cavity that we use for our analysis. In the next
section we describe the experimental arrangement and the relevant
Bell-type inequality~(BI). We wish to emphasize that although the
BI used by us is the same as in \cite{rydberg2}, the cavity dynamics
considered by us (the steady-state dynamics is discussed in Section III) differs crucially form the one used in
\cite{rydberg2}.  We are hence able to take into account micromaser
dissipation in a more realistic manner, and also analyse the
microlaser by incorporating atomic decay. In Section IV we present and discuss
our results. Section V is reserved for some concluding remarks.

\section{Violations of Bell's inequality in a microcavity}

We consider the following experimental scenario. 
A two-level atom  initially in its upper excited state $|e>$ traverses
a high-Q single mode cavity.  
The cavity is in a steady state
and tuned to a single mode resonant with the transition $|e> \to |g>$. 
The emerging
single-atom wavefunction is a superposition of the upper $|e>$ and
lower $|g>$ state, and it leaves an imprint on the photonic wavefunction
in the cavity. After leaving the cavity, the atom passes through
an electromagnetic field which gives it a $\pi/2$ pulse  the
phase of which can be varied for different atoms. The atom  then reaches the
detector, placed at a sufficient distance, capable of detecting the atom only
in the upper or lower state. Thus, the role of the $\pi/2$ pulse may be
considered as a component of the detection mechanism in the 
experiment\cite{pulse}.
During the whole process, dissipation takes place, and is taken into account.
Next, this process is repeated for a similar second atom. The important difference is
that the second atom interacts with a photonic wavefunction which has been modified
due to the passage of the first atom. There is no direct interaction between the
two atoms, although secondary correlations develop between them. In other words,
though there is no spatial overlap between the two atoms, the entanglement of
their wavefunctions with the cavity photons can be used to formulate local-realist
bounds on the detection probabilities for the two atoms\cite{rydberg1,rydberg2}.

The interplay of the atomic statistics with the photonic statistics plays a 
crucial role in the investigation of nonlocality here. The initial state of the
cavity is built up by the passage of a large number of atoms, but only
one at a time, through it. The
pump parameter and the atom-photon interaction time are key inputs for the profile
of the resultant photonic wavefunction which in turn governs the nature of
entanglement between  two successive experimental atoms detected in their upper
or lower states by the detector. As stated earlier, dissipation due to the interaction
of the pumping atoms with their reservoir, as well as the loss of cavity photons can
be controlled, and their effects on the statistics of detected atoms can be studied.
The formalism used by us has another generic feature. The effects of decoherence on
nonlocality can be studied in context of the micromaser, as well as the 
microlaser, its optical counterpart.

It is easy to obtain a Bell-type inequality suitable for the scenario considered
by us in analogy to Bell's original reasoning. Two level Rydberg atoms are analogous
to spin-$1/2$ systems, and the phase of the electromagnetic field plays the role of
the polarization axis of the Stern-Gerlach apparatus used for spin-$1/2$ systems.
In fact, several local realist bounds have earlier been derived to tailor such a 
situation\cite{rydberg1,rydberg2}. Let us very briefly describe one 
such derivation\cite{rydberg2}
which we shall use in the present analysis.
Assigning the  value $+1$ for the atom  detected in the upper
state $|e>$, and $-1$ for the lower state $|g>$, one can in any local realist theory
define functions $f(\phi_1) = \pm 1; \,  g(\phi_2) = \pm 1$ 
describing the outcome of measurement on the atom 1 and 2 when the phase of the 
electromagnetic field giving $\pi/2$  pulse to the atoms
is set to be $\phi_1$ and $\phi_2$ for the respective atoms. The ensemble
average for double click events is therefore defined as
\begin{equation}
E^{\l}(\phi_1,\phi_2) = \int d\l f(\phi_1)g(\phi_2)
\end{equation}
where $\l$ is a suitable probability measure on the space of all possible states.

The quantum mechanical expectation value for double click events is calculated from
the probabilities of all possible double-click sequences. This is given by
\begin{eqnarray}
E(\phi_1,\phi_2) &=& P_{ee}(\phi_1,\phi_2) + P_{gg}(\phi_1,\phi_2) \nonumber \\
&-& P_{eg}(\phi_1,\phi_2)
- P_{ge}(\phi_1,\phi_2)
\end{eqnarray}
where $P_{eg}(\phi_1,\phi_2)$ stands for the probability that the first atom is
found to be in state $|e>$ after traversing the $\pi/2$ pulse with phase $\phi_1$,
and the second atom is found to be in state $|g>$ with the phase of the $\pi/2$ 
pulse being $\phi_2$ for its case. Defining $E_0 = E^{\l}(\phi_1=\phi_2)$ and
$M_0 = P_{ee}(\phi_1=\phi_2) + P_{gg}(\phi_1=\phi_2)$, and assuming perfect
detections, 
it follows that
$E_0 = 2M_0 - 1$.
Further, it is easy to see that $f(\phi) = +g(\phi)$ 
with probability $M_0$, and $f(\phi) = -g(\phi)$
with probability $(1-M_0)$. Hence, $E^{\l}(\phi_1,\phi_2)$ can be written as
\begin{eqnarray}
E^{\l}(\phi_1,\phi_2) = 
 E_0\int d\lambda f(\phi_1)f(\phi_2)
\end{eqnarray}
Now, one can define a Bell sum
\begin{eqnarray}
B &\equiv& |E^{\l}(\phi_1,\phi_2) - E^{\l}(\phi_1,\phi_3)| \nonumber \\
&+&
sign(E_0)[E^{\l}(\phi_2,\phi_3) - E_0] 
\end{eqnarray}
It follows immediately from (2-4) that $B \le 0$. In the next section we shall calculate this Bell
sum $B$ and see how it evolves for various values of the cavity parameters. 
It is convenient to set the values of the phases $\phi_1=0$, $\phi_2=\pi/3$,
and $\phi_3=2\pi/3$, as for these values the Bell-type inequality is always violated, i.e., $B > 0$
for the case of an idealised micromaser\cite{rydberg2}.

\section{Steady-state micromaser/microlaser photon statistics}

In realistic situations, one must consider the interacting systems (atoms
as well as cavity field) coupled to their respective reservoirs. 
The couplings
are governed by their equations of motion\cite{book},

\begin{eqnarray}
\dot{\rho}\vert_{atom-reservoir} &=& - \gamma (1+\bar{n}_{th}) (s^{+}s^{-}\rho
-2s^{-}\rho s^{+} + \rho s^{+}s^{-}) \nonumber \\
 &-& \gamma \bar{n}_{th} (s^{-}s^{+} \rho
-2s^{+}\rho s^{-} + \rho s^{-}s^{+})
\end{eqnarray}
for the atom and
\begin{eqnarray}
\dot{\rho}\vert_{field-reservoir} &=& - \kappa (1+\bar{n}_{th}) (a^{\dg}a \rho  
- 2a\rho a^{\dg} + \rho a^{\dg} a ) \nonumber \\
&-& \kappa \bar{n}_{th} (a a^{\dg} \rho
-2 a^{\dg}\rho a + \rho a a^{\dg})
\end{eqnarray}
for the field. $\rho$ is the reduced density operator obtained after tracing
over the reservoir. $\gamma$ and $\kappa$ are the decay constants for the
atom and the field respectively. $\bar{n}_{th}$ is the average photon
number representing the reservoir. $s^{+}$ and $s^{-}$ are the usual
Pauli operators for the pseudo-spin representation of the two-level model
of the atom. $a(a^{\dg})$ is the photon annihilation (creation) operator.

The dynamics we are interested in, involves two-level atoms steamed into a 
single mode cavity in such a way that there is at most one atom in the cavity
at any time.  Thus we have sequences of events (atom-field interactions)
taking place randomly, but with each event of a fixed duration $\tau$. This
interaction is governed by 
\begin{equation}
\dot{\rho}\vert_{atom-field} = -i [H, \rho]
\end{equation}
where $H = g(s^{+}a + s^{-}a^{\dg})$ is the well known 
Jaynes-Cummings\cite{jaynes} Hamiltonian with $g$ being the coupling constant.

Thus, we have to solve the equation of motion
\begin{equation}
\dot{\rho} = \dot{\rho}\vert_{atom-reservoir} + \dot{\rho}\vert_{field-reservoir} 
+ \dot{\rho}\vert_{atom-field}
\end{equation}
where the terms on the r.h.s. are the r.h.s' of (5), (6) and (7) respectively.
For the duration between two events, we have to solve the equation of motion
(6) only.

The steady-state photon statistics of the cavity field undergoing such
dynamics has been derived in\cite{maser1}. In the cavity photon number 
representation, the probabilities $P_n = <n|\rho_f^{(ss)}|n>$ are given by
\begin{equation}
P_{n}=P_{0}\prod_{m=1}^{n}v_{m}
\end{equation}
$P_{0}$ is obtained from the normalisation $\sum^{\infty}_{n=0}P_{n}=1$.
The $v_{n}$ are given by the continued fractions
\begin{equation}
v_n=f^{(n)}_{3}/(f^{(n)}_{2} + f^{(n)}_{1}v_{n+1})
\end{equation}
with $f^{(n)}_{1}=(Z_{n}+C_{n})/\kappa$,
$f^{(n)}_{2}=-2N+(Y_{n}+B_{n})/\kappa$
and $f^{(n)}_{3}=-(X_{n}+A_{n})/\kappa$. $\kappa$ is the cavity bandwidth
and $N=R/2\kappa$ is the number of atoms passing through the cavity in a
photon lifetime. $A_{n}=2n\kappa\bar{n}_{th}$,
$B_{n}=-2\kappa(n+\bar{n}_{th}+2n\bar{n}_{th})$ and
$C_{n}=2(n+1)(\bar{n}_{th}+1)\kappa$. $X_{n}$, $Y_{n}$ and $Z_{n}$
are given by
\begin{equation}
X_{n}=R\sin^{2}(g\sqrt{n}\tau)\exp\{-[\gamma+(2n-1)\kappa]\tau\}
\end{equation}
\begin{eqnarray}
Y_{n} = \frac{1}{2}R\Bigl(\{2\cos^2[g\sqrt{n+1}\tau]-\frac{1}{2}
(\gamma/\kappa+2n+1) \nonumber \\
+F_{1}(n-1)\}\exp\{-[\gamma
 +(2n+1)\kappa]\tau\}+
[\frac{1}{2}(\gamma/\kappa+2n+1) \nonumber \\
-F_{2}(n-1)]\exp\{-[\gamma+
(2n-1)\kappa]\tau\}\Bigr),
\nonumber
\end{eqnarray}
and
\begin{eqnarray}
Z_{n} & = & 
\frac{1}{2}R\Bigl([\frac{1}{2}(\gamma/\kappa+2n+3)+F_{2}(n)]
\exp\{-[\gamma+(2n+1)\kappa]\tau\}
\nonumber \\
 -[\frac{1}{2}(\gamma/\kappa+2n+3)+F_{1}(n)]\exp\{-[\gamma
+(2n+3)\kappa]\tau\}\Bigr)
\nonumber
\end{eqnarray}
 $\gamma$ represents reservoir induced spontaneous
emission from the upper to the lower masing level.
The functions $F_i$ are
\begin{eqnarray}
F_{i}(n) = 
\frac{\kappa/4g}{(\sqrt{n+2}-\sqrt{n+1})^2} \nonumber \\
\Biggl[\frac{\gamma}{\kappa}
(\sqrt{n+2}-\sqrt{n+1}) \sin{(2g\sqrt{m}\tau)} 
 -\frac{\gamma}{g}\cos{(2g\sqrt{m}\tau)} \nonumber \\
-[2n+3+2\sqrt{(n+1)(n+2)}](\sqrt{n+2}-\sqrt{n+1})\sin{(2g\sqrt{m}\tau)}\Biggr]
\nonumber \\
 +\frac{\kappa/4g}{(\sqrt{n+2}+\sqrt{n+1})^{2}} \nonumber \\
\Biggl[\pm\frac{\gamma}{\kappa}(\sqrt{n+2}+\sqrt{n+1})\sin{(2g\sqrt{m}\tau)} -\frac{\gamma}{g}\cos{(2g\sqrt{m}\tau)}
\nonumber \\
 \mp[2n+3-2\sqrt{(n+1)(n+2)}](\sqrt{n+
2}+\sqrt{n+1})\sin{(2g\sqrt{m}\tau)}\Biggr]
\end{eqnarray}
where $m=n+2$ and $n+1$ for $i=1$ and $2$, respectively, with the
upper sign for $i=1$.

 The experimental atoms on
which we plan to test the Bell's inequality (BI), encounter this steady state
radiation field $\rho_f^{(ss)}$. The atom-field interaction is, as
mentioned earlier governed by the Jaynes-Cummings Hamiltonian
$H$. After interaction with the cavity, the first experimental atom
emerges in a superposition of the upper ($|e>$) and the lower
($|g>$) states and experiences a $\pi/2$ pulse with phase
$\phi_1$. The probability of detection of the atom in the upper
state $|e>$ and the lower state $|g>$ can be written respectively as
\begin{eqnarray}
P_e = {\rm Tr._f}{\cal P}_e \nonumber \\
P_g = {\rm Tr._f}{\cal P}_g
\end{eqnarray}
with 
\begin{eqnarray}
{\cal P}_e = {1 \over 2}[{\cal A}\rof {\cal A}^{\dg} + {\cal D}\rof
{\cal D}^{\dg} - \{e^{-i\phi_1}{\cal A}\rof {\cal D}^{\dg} +
e^{i\phi_1}{\cal D}\rof {\cal A}^{\dg}\}] \nonumber \\
{\cal P}_g = {1 \over 2}[{\cal A}\rof {\cal A}^{\dg} + {\cal D}\rof
{\cal D}^{\dg} + \{e^{-i\phi_1}{\cal A}\rof {\cal D}^{\dg} +
e^{i\phi_1}{\cal D}\rof {\cal A}^{\dg}\}] 
\end{eqnarray} 
where trace is taken over the cavity field and the operators ${\cal
A}$ and ${\cal D}$ are given by
\begin{eqnarray}
{\cal A} = {\rm cos}(gt\sqrt{a^{\dg}a + 1}) \nonumber \\
{\cal D} = -ia^{\dg}{{\rm sin}(gt\sqrt{a^{\dg}a + 1}) \over
\sqrt{a^{\dg}a + 1}}
\end{eqnarray}
After the passage of the first atom through the cavity and its
detection in, for example, the state $|e>$, the second atom encounters
the cavity field with density operator $\rho_f^{(2)}$ given by
\begin{equation}
\rho_f^{(2)} = [{\cal A}\rof {\cal A}^{\dg} + {\cal D}\rof
{\cal D}^{\dg} - \{e^{-i\phi_1}{\cal A}\rof {\cal D}^{\dg} +
e^{i\phi_1}{\cal D}\rof {\cal A}^{\dg}\}]
\end{equation}
(since ${\rm Tr._f}{\cal P}_e = 1/2$).
The phase of the $\pi/2$ pulse is set to $\phi_2$ for the second
atom. ${\cal P}_e$ for the second atom is given by
\begin{equation}
{\cal P}_e^{(2)} =  {1 \over 2}[{\cal A}\rho_f^{(2)} {\cal A}^{\dg} + {\cal D}\rho_f^{(2)}
{\cal D}^{\dg} - \{e^{-i\phi_1}{\cal A}\rho_f^{(2)} {\cal D}^{\dg} +
e^{i\phi_1}{\cal D}\rho_f^{(2)} {\cal A}^{\dg}\}]
\end{equation}
The conditional probability $P_{ee}(\phi_1,\phi_2)$ is thus given
by 
\begin{equation}
P_{ee}(\phi_1,\phi_2) = {\rm Tr._f}{\cal P}_e^{(2)}
\end{equation}
$P_{gg}$ is obtained similarly, and using the relations
$E(\phi_1,\phi_2) = 2P(\phi_1,\phi_2) - 1$ and $P(\phi_1,\phi_2) =
P_{ee}(\phi_1,\phi_2) +  P_{gg}(\phi_1,\phi_2)$ one obtains the
values of $E$ in the Bell sum (4). We perform this exercise
numerically with the values of the phases in the Bell sum (4) set
to $\phi_1=0$, $\phi_2=\pi/3$, and $\phi_3= 2\pi/3$.  
 
It will be appropriate to mention here that although 
decoherence
effects are inherent in the build up of the cavity field to its steady
state since it encounters a large number of atoms over the time required
for the steady state to be reached, the dynamics of a single experimental
atom interacting with this field for a short duration will be unaffected
by the decoherence effects, as has been observed in the cavity-QED of
Jaynes-Cummings interaction\cite{cqed}. It was shown there
that typical durations
of atom-field interactions in realistic environments can be as large as
$t \sim 10/g$ up to which decoherence effects are insignificant. We again stress
here that even though the individual atom-field interaction time in the 
dynamics of the cavity field, pumped repeatedly with atoms, is 
uniformly of this order, the dissipative forces there play a crucial role 
due to the number of atoms ($\gg 1$) and the time ($\gg 10/g$) involved
in reaching the steady state. The same argument applies to the case
of atomic damping, as well. Indeed, we
keep the above argument in mind in our choice of parameters while
calculating the probabilities in the Bell sum (4).

The micromaser model considered by us differs crucially from that in
\cite{rydberg2} because in the latter model cavity dissipation is neglected
whenever an atom is present in the cavity. Our model takes
into account dissipation even during
the short atom-field interaction times. This {\it a priori} small effect
on cavity photon statistics gets magnified due to the requirement of
passage of a large
number of atoms through the cavity for it to reach its steady state.
Thus, as expected, the resultant steady-state photon statistics in our
model is clearly different  from
the one used in \cite{rydberg2}. Let us again emphasize, that although the Bell's inequality (BI) we  propose to
test is the same as in \cite{rydberg2}, our cavity dynamics are fundamentally
different: (a) our micromaser model (discussed in detail in \cite{maser1})
is more realistic, and (b) we are also
able to analyse the microlaser by incorporating atomic decay.

\section{Effects of cavity dissipation and atomic decay on the Bell
sum}

In the previous section we have presented steady-state cavity
dynamics describing both the micromaser as well as the microlaser
within a unified framework. However, their distinctive features are
manifested in the choice of parameters which we use below to study the
violation of BI in both separately. Our choice of different sets of
experimentally attainable parameters for both the micromaser 
and the microlaser is motivated by
the following considerations. Recall that the experimental atoms
are pumped into the cavity with steady-state photon statistics at 
such a rate that at most one atom
is present in the cavity at any time. So, there are two situations: (a) atom
present in the cavity, and (b) cavity empty of atom. Whatever be the situation,
cavity dissipation, i.e., the interaction of the cavity mode with its
reservoir continues uninterrupted. This process is of crucial importance
in micromaser dynamics, and has been discussed in detail in
\cite{maser1}. However, while considering micromaser dynamics, 
one can safely set atomic decay to zero. This is because the
Rydberg levels involved in the micromaser have a lifetime of about $0.2s$, whereas the atomic
flight time through the cavity (atom-field interaction time) is typically
$35\mu s$. We present the variation of the Bell sum $B$ with
respect to the  parameter $D$ (which is nothing but the Rabi angle
$\phi$ scaled by the pump rate $N$, i.e., $D=\phi\sqrt{N}$) for the case 
of the micromaser in Figure 1.

\begin{figure}
\centerline{\epsfig{file=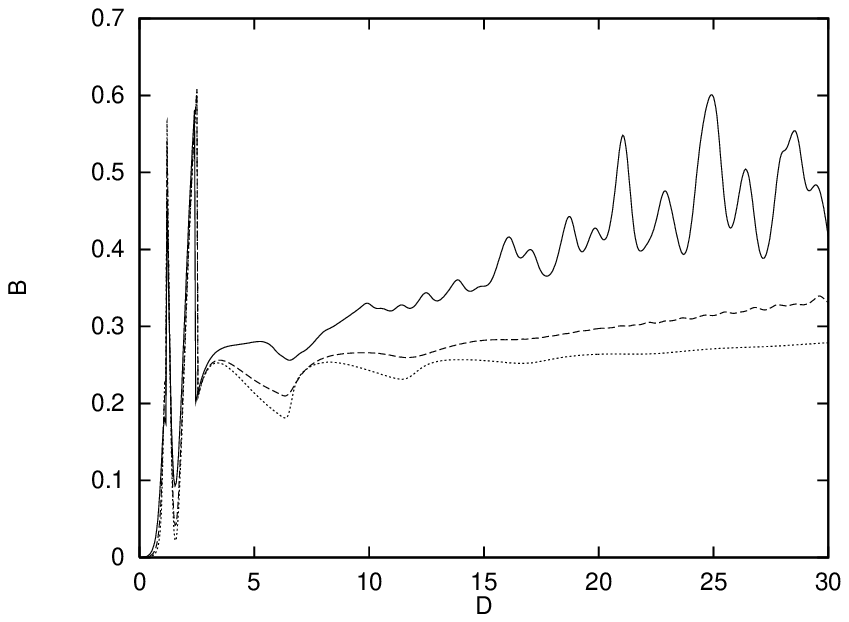,height=3.0in}}
\caption\protect{
Violation of Bell's inequality in a micromaser\cite{maserdata}. Atoms
in the
upper of the two Rydberg levels are streamed into a cavity, one at a time,
in such a way that
the flight time of an atom $\tau$ is much shorter than the lifetime of the
long-lived Rydberg levels. Hence, we set the atomic decay constant $\gamma = 0$.
The average thermal photons $\bar{n}_{th} = 0.15$ in
the cavity represent its temperature at $0.5 K$. The leakage of the cavity
photons is represented by $\kappa/g = 0.1\times 10^{-6}$.
The pump rate $N$, the number of
individual atoms that pass through the cavity in a photon lifetime, $= 20$
(full), 50 (broken), and 100 (dotted). $D = \phi\sqrt{N}$ where the Rabi 
angle $\phi =g\tau$. The parameters
are very close to the experimental data in \cite{maserdata}.}
\end{figure}

Our results show clearly that the Bell sum
reflecting the degree of nonlocality exhibited in the atom-atom secondary
correlations depends heavily on cavity dissipation. In particular,
it is seen that the value of Bell sum $B$ decreases with the increase of
pump rate $N$ for a large range of interaction times $\tau$. This can be 
understood from the way dissipative effects creep
into the dynamics through two parameters $N$ and $\tau$\cite{maser1}. 
The genesis of atom-photon and the resultant atom-atom entanglement
competes with decoherence in an interesting fashion over time. 
For shorter values of single atom interaction times we find that the
correlations build up sharply with $\tau$, and the peak value of $B$ 
signifying maximum violation of BI is larger for higher values of $N$, 
(as a magnification of Figure 1 reveals). The maximum violation (the value
of $B$ at its second peak) is $0.5812$ for $N=20$ (full line), $0.6079$
for $N=50$ (broken) and $0.6241$ for $N=100$ (dotted) (See Figure 2 where we 
magnify the curves in the region $D\le 5$ of Figure 1 for a clear display of 
these peaks).
This feature is a curious example of multiparticle induced nonlocality.
It is analogous to the enhancement of nonlocality for multiparticle systems,
and is in conformity with the mathematical demonstration of larger violation of
BI with increase in the number of particles 
involved\cite{macro,larjn}. For
short interaction times, naturally the effects of decoherence are too small
to affect the correlations. One noticeable feature  in Figure 1 is the
structures for low values of $N$ (full line). This originates from ``trapped
state'' dynamics of photonic statistics\cite{maser1,trapped}  where it has been shown that dissipative effects
gradually wash out such states, as can be seen from the broken and
dotted lines. 

\begin{figure}
\begin{center}
\centerline{\epsfig{file=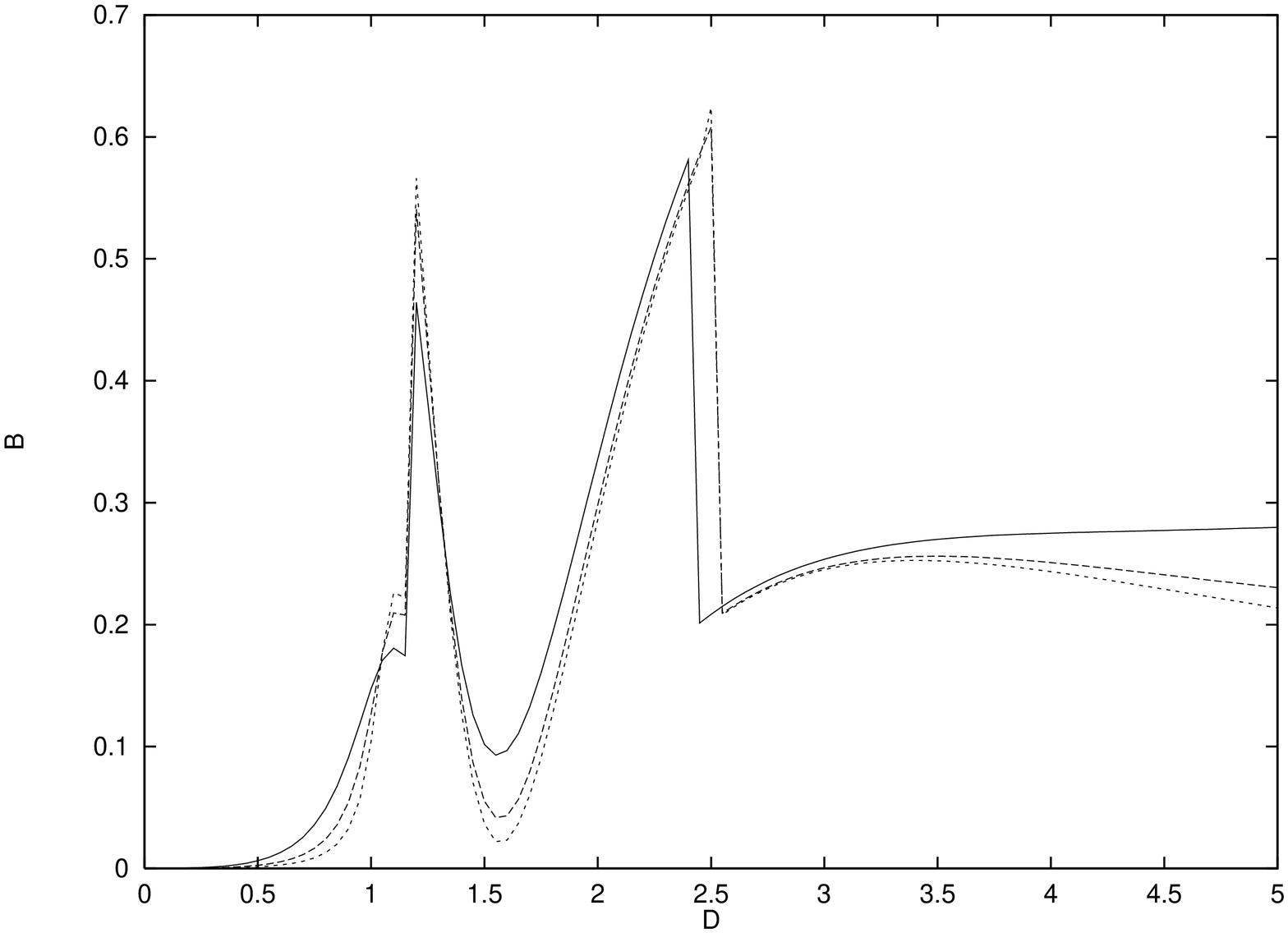,height=4.0in,angle=270}}
\caption{Enhancement of nonlocality for increased $N$ in the
micromaser. The maximum value of $B$ is 0.5812 for $N=20$ (full
line), 0.6079 for $N=50$ (broken), and 0.6241 for $N=100$
(dotted).}
\end{center}
\end{figure}

Let us now consider the violation of BI in the microlaser. 
It is well known
that atomic decay is an important factor in the microlaser where atomic
levels at optical frequencies are involved. Although decay is
unimportant for the dynamics of a single atom interacting with the field
up to a certain interaction time,
its accumulated effect for a large number of atoms is crucial for the
evolution of the microlaser field to a steady state.
Furthermore, the interaction
time of the atom with the $\pi/2$ pulse (in this case it is $gt=\pi/2$)
is far less compared to even individual atom-field interaction times of
interest in microlaser dynamics. For this reason
the effect of atomic decay can be neglected during  interaction
of the atom with the $\pi/2$ pulse. Our results for the
microlaser are shown in Figure 3.

\begin{figure}
\centerline{\epsfig{file=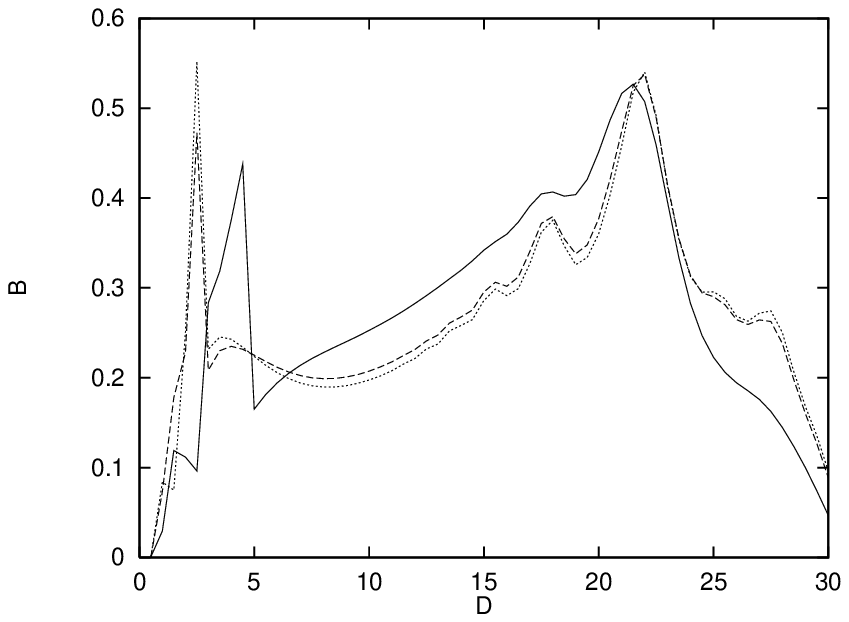,height=3.0in}}
\caption\protect{
Demonstration of nonlocality in a microlaser, the optical
counterpart of the micromaser. The results of our numerical simulations
can be tested in a microlaser of the type in \cite{microlaser}.
Atomic levels having transition frequency in the optical regime have shorter
lifetimes compared to  Rydberg levels, and hence, we set $\gamma/g = 0.1$.
At optical frequencies, thermal photons are very close to zero, and thus we
take $\bar{n}_{th} = 0$. $N=100$.
The cavity leakage rate is $\kappa/g = 0.01$ (full), 0.001 (broken) and
0.0001 (dotted).}
\end{figure}

The effect of decoherence (cavity leakage rate $\kappa/g$) on the 
violation of BI is clearly seen in the microlaser.
One can check, similar to the case of the micromaser that
 the value of Bell sum $B$ decreases with the increase of
pump rate $N$ for a large range of interaction times $\tau$.
But in
case of the microlaser, atomic damping $\gamma$ is a dominating factor, in
contrast to the cavity photon loss in the micromaser. This 
makes the Bell sum fall off rapidly for large values of $\tau$.
 The second peak in $B$ however, is a consequence of the Jaynes-Cummings
dynamics and  survives such dissipation\cite{maser2}.

\section{Conclusions}

To summarize, we have shown that a demonstration of nonlocality, encompassing
several of its varied aspects
in atom-photon interactions in cavities and the effects of decoherence on it,
can be possible in experimental set-ups already in operating conditions for
the micromaser\cite{maserdata}, as well as for the 
microlaser\cite{microlaser}. The formalism chosen enables us to consider
the steady-state dynamics of both the micromaser as well as the
microlaser in the presence of atomic decay and cavity dissipation 
within a unified framework. Their distinguishing
features are brought about by the different values of the
parameters chosen to analyze the violation of BI in them separately.
Certain notable
features, such as enhancement of nonlocality for increased number of atoms,
when decoherence effects are small,
can  be observed. We have seen how such features can be quantitatively
monitored by control of decoherence. In an actual experiment, certain points
have to be borne in mind. A few atoms may go undetected between two detector
clicks. However, the steady state nature of the cavity field contributes to
making the effect of this on the Bell sum insignificant compared to the 
effect of decoherence which we have probed in detail. Finally, the observed 
magnitude
of violation of BI would be brought down by 
finite detector efficiency. Nevertheless, our selection of the particular type
of BI\cite{rydberg2}, and the phases of $\phi$, ensure that this BI is always violated
for the range of parameters chosen irrespective of the efficiency factor of
the detector, which can in any case be accounted for easily by the introduction
of appropriate scaling factors in the expressions of the various probabilities
appearing in the Bell sum.

 
\end{document}